\documentclass[english]{article}
\usepackage{amsfonts}
\usepackage{amsmath}
\usepackage{amssymb}
\usepackage{amsthm}
\usepackage{verbatim}
\usepackage{float}
\usepackage{subfig}
\usepackage{multirow}
\usepackage[colorlinks,citecolor=blue,urlcolor=blue,linkcolor=red]{hyperref}
\usepackage{fullpage}
\usepackage[affil-it]{authblk}
\usepackage{siunitx}

\usepackage{textcomp}
 \usepackage{tabularx}
\usepackage{booktabs}
\usepackage{caption}
\usepackage{graphicx}
\usepackage{libertine} 

\renewcommand{\phi}{\varphi}
\renewcommand{\epsilon}{\varepsilon}

\begin{document}
\title{DeshadowGAN: A Deep Learning Approach to Remove Shadows from Optical Coherence Tomography Images}
%\date{}

\author[1]{Haris Cheong}
\author[1]{Sripad Krishna Devalla}
\author[1]{Tan Hung Pham}
\author[1]{Zhang Liang}
\author[2]{Tin Aung Tun}
\author[2]{Xiaofei Wang}
\author[2, 3]{Shamira Perera}
\author[2, 3, 4, 6, 7]{Leopold Schmetterer}
\author[1, 2]{Aung Tin}
\author[1,8,9]{Craig Boote}
\author[4 $\star$]{Alexandre H. Thi\'{e}ry}
\author[1,2 $\star$]{Micha\"el J. A. Girard}

\affil[1]{Ophthalmic Engineering and Innovation Laboratory, Department of Biomedical Engineering, Faculty of Engineering, National University of Singapore, Singapore.}
\affil[2]{Singapore Eye Research Institute, Singapore National Eye Centre, Singapore.}
\affil[3]{Duke-NUS Graduate Medical School, Singapore}
\affil[4]{Department of Statistics and Applied Probability, National University of Singapore, Singapore.}
\affil[5]{Beijing Advanced Innovation Center for Biomedical Engineering, School of Biological Science and Medical Engineering, Beihang University, Beijing, China}
\affil[6]{Nanyang Technological University, Singapore}
\affil[7]{Department of Clinical Pharmacology, Medical University of Vienna, Austria.}
\affil[8]{School of Optometry and Vision Sciences, Cardiff University, UK.}
\affil[9]{Newcastle Research and Innovation Institute, Singapore.}

\affil[$\star$]{Both authors contributed equally and are both corresponding authors.}
\affil[$\star$]{email:mgirard@invivobiomechanics.com, a.h.thiery@nus.edu.sg}

\maketitle

%\tableofcontents

%
% -------------------------------------------------------------------------------------------
% ABSTRACT
% -------------------------------------------------------------------------------------------
%
\begin{abstract}
\noindent
{\bf Purpose:}
To remove retinal shadows from optical coherence tomography (OCT) images of the optic nerve head (ONH).\\
\noindent
{\bf Methods:}
2328 OCT images acquired through the center of the ONH using a Spectralis OCT machine for both eyes of 13 subjects were used to train a generative adversarial network (GAN) using a custom loss function.Image quality was assessed qualitatively (for artifacts) and quantitatively using the intralayer contrast – a measure of shadow visibility ranging from 0 (shadow-free) to 1 (strong shadow) and compared to compensated images. This was computed in the Retinal Nerve Fiber Layer (RNFL), the Inner Plexiform Layer (IPL), the Photoreceptor layer (PR) and the Retinal Pigment Epithelium (RPE) layers. \\
\noindent
{\bf Results:}
Output images had improved intralayer contrast in all ONH tissue layers. On average the intralayer contrast decreased by $33.7 \pm 6.81\%$, $28.8 \pm 10.4\%$, $35.9 \pm 13.0\%$, and $43.0 \pm 19.5\%$ for the RNFL, IPL, PR, and RPE layers respectively, indicating successful shadow removal across all depths. This compared to $70.3 \pm 22.7\%$, $33.9 \pm 11.5\%$, $47.0 \pm 11.2\%$, $26.7 \pm 19.0\%$ for compensation.   Output images were also free from artifacts commonly observed with compensation.\\
\noindent
{\bf Conclusions:}
DeshadowGAN significantly corrected blood vessel shadows in OCT images of the ONH. Our algorithm may be considered as a pre-processing step to improve the performance of a wide range of algorithms including those currently being used for OCT image segmentation, denoising, and classification.\\
\noindent
{\bf Translational Relevance:} DeshadowGAN could be integrated to existing OCT devices to improve the diagnosis and prognosis of ocular pathologies. 
\end{abstract}

\section{Introduction}
\label{sec:intro}
Glaucoma is a leading cause of irreversible blindness and occurs due to the death of retinal ganglion cells.\cite{RN1} In its most common form, there are no symptoms, making regular diagnostic tests crucial for early detection and treatment.\cite{RN2} Clinical research suggests that glaucomatous eyes have a unique biomechanical profile that allows differentiation from non-glaucomatous eyes.\cite{RN3}\\

Optical coherence tomography (OCT) has been proven to be a promising tool for automated classifiers to identify glaucomatous eyes from healthy eyes.\cite{RN4} It uses low-coherence light to capture micrometer resolution, three-dimensional images, allowing in-vivo visualization of a patient’s retinal layers, making it the least invasive choice among other diagnostic tools as it does not require contact with the eye or eye drops to be applied before testing.\cite{RN5}  \\

However, due to the high absorption scattering property  of retinal vessels, information from locations beneath these vessels are is significantly decreased.\cite{RN6} This causes artifacts known as retinal shadows which decrease the readability of retinal OCT B-scans. These artifacts also result in errors in thickness assessment of the RNFL, which has clinical implications for the diagnosis management of glaucoma where changes in the RNFL need to be monitored accurately over time.\cite{RN7} The shadows are also the main artifact occluding deep structures such as the lamina cribrosa (LC).\cite{RN8} Consequently, it is crucial to develop algorithms to replenish the information lost within these shadows to achieve the best automated retinal layer segmentation and eventual diagnosis accuracy.\\

Fabritius et al.\cite{RN9} described a compensatory method to reduce the effects of vessel artifacts on interpretation of the retinal pigment epithelium (RPE) layer. Girard et al.\cite{RN10} also improved the quality of OCT images through compensation methods by correcting the effects of light attenuation and by better estimating the optical properties (i.e. reflectivity) of the tissues. These predictions are, however, only estimations or based on simple optical models that may result in secondary artifacts being produced such as inverted shadows.\\

Artificial intelligence techniques have been extensively applied to shadow removal algorithms for normal images with varying levels of success.\cite{RN64,RN65} The landmark article \cite{RN66} introduced  generative adversarial networks (GANs) architectures. This technique paved the way for GANs to be applied for other purposes, such as shadow removal,\cite{RN64} shadow detection,\cite{RN68,RN69} and unwanted artifact removal.\cite{RN70}\\

In this study, we aimed to test whether a custom GAN (referred herein as DeshadowGAN) could automatically detect and remove shadows according to a predicted {\it shadow score} in order to improve the quality of OCT images of the ONH.

\section{Methods}
\label{sec:methods}
\subsection{Patient Recruitment}
A total of $13$ healthy subjects were recruited at the Singapore National Eye Centre. All subjects gave written informed consent. This study adhered to the tenets of the Declaration of Helsinki and was approved by the institutional review board of the hospital. The inclusion criteria for healthy subjects were: an intraocular pressure (IOP) less than $21$ mmHg, and healthy optic nerves with a vertical cup-disc ratio (VCDR) less than or equal to $0.5$.

\subsection{Optical Coherence Tomography Imaging}
Recruited subjects were seated and imaged in dark room conditions by a single operator. A standard spectral domain OCT (Spectralis, Heidelberg, Germany) system was used to image both eyes of each subject. We obtained 97 horizontal B-scans ($32 \, \mu m$ distance between B-scans; 384 A-scans per B-scan) from a rectangular area of $\ang{15} \times \ang{10}$ centered on the ONH. $75$ times signal-averaged images were obtained from multi-frame volumes. In total, our training set consisted of $2,328$ multiframe baseline B-scans from $24$ 3D volumes. Our test set consisted of $291$ multiframe baseline B-scans from three 3D volumes. 

\subsection{DeshadowGAN: Overall description}
Our algorithm comprised of two networks competing with each other. One network, called the shadow detection network, predicted which pixels were considered as shadowed pixels. The other network, called the shadow removal network, aimed to prevent the shadow detection network from finding shadowed pixels by removing shadows from baseline B-scans. \\

\begin{figure}[H] 
    \centering
    \includegraphics[width=\textwidth,height=\textheight,keepaspectratio]{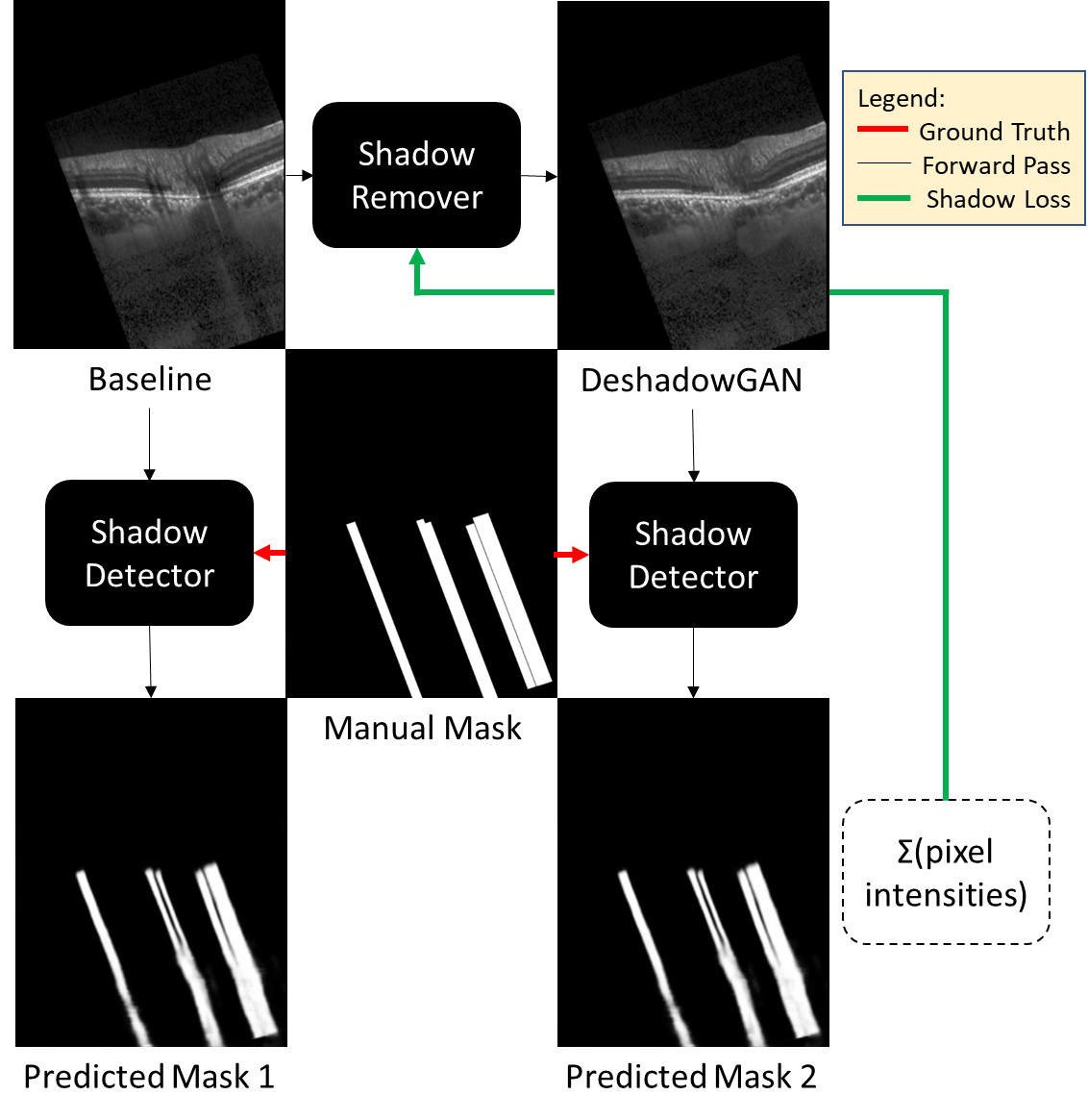}
    \caption{Overall algorithm training diagram.}
    \label{fig:1}
\end{figure}

First, we trained the shadow detection network five times on baseline images with their corresponding manually segmented shadow mask as the ground truth. Binary segmentation masks (size: $496 \times 384$) were manually created for all $2,328$ B-scans using ImageJ\cite{RN12} by one observer [HC]. Next, we trained the shadow removal network once by passing the baseline as input and using the predicted binary masks as part of the loss function. Lastly, we trained the shadow detector network another five times with the output from the shadow removal network, and another five times with the manually segmented binary masks as ground truth. (Figure \ref{fig:1})

\subsubsection{Shadow Detection Network}
\begin{figure}[H] 
    \centering
    \includegraphics[width=\textwidth,height=\textheight,keepaspectratio]{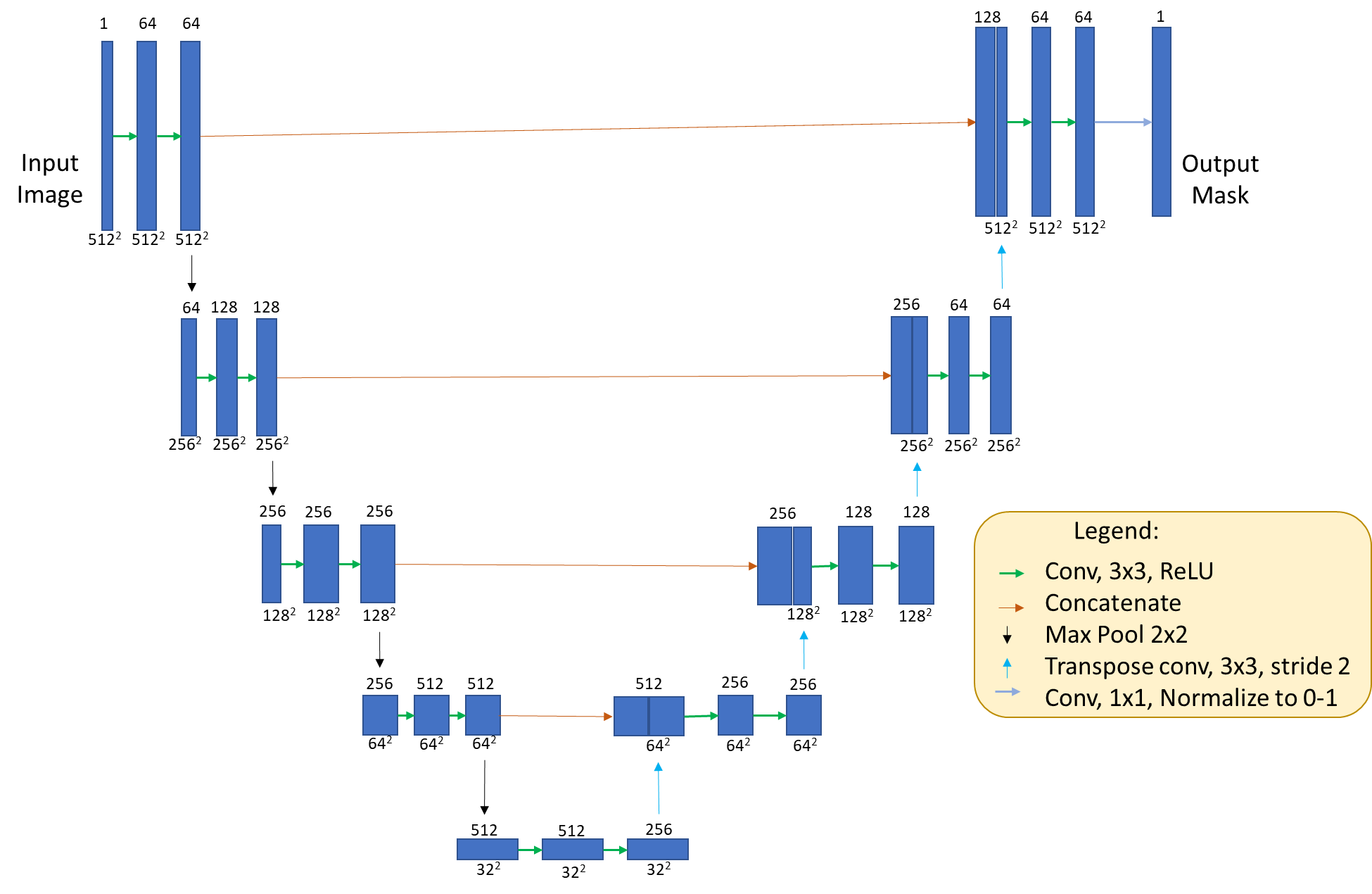}
    \caption{Shadow Detector network architecture. Numbers on top of each rectangle represents the number of feature maps and numbers below each rectangle represents the feature map size. Network consists of 13.4M parameters, occupying $648$ Mo of RAM on a single Nvidia GTX1080Ti GPU.}
    \label{fig:2}
\end{figure}

A neural network inspired by the U-Net architecture\cite{RN13} (Figure \ref{fig:2}) was trained with a simple Binary Cross Entropy loss using the hand-crafted segmentation masks as ground truth. This network had a sigmoid layer as its final activation, making it a per-pixel binary classifier. It was then trained with original images concatenated with the output from a shadow removal network, using the manually segmented masks as ground truth.

The shadow detection network first performed two convolutions with kernel size $3$ and stride $1$, followed by a ReLU activation\cite{RN71} after each convolution. Then, images were downsampled using a $2 \times 2$ maxpooling operation, halving the size of the height and width of the feature maps. This occurred $4$ times, with the number of feature maps at each smaller size increasing from $1$ to $64$, $128$, $256$, $512$ respectively. \\

The shadow detection network comprised of two towers. A downsampling tower at each stage sequentially halved the dimensions of the baseline image (size: $512 \times 512$) via maxpooling to capture the contextual information (i.e. spatial arrangement of tissues), and an upsampling tower which sequentially restored it back to its original resolution to capture the local information (i.e. tissue texture).\cite{RN104} Transposed convolution was performed $4$ times in the upsampling tower for the predicted segmentation masks to be size $512 \times 512$, before passing to a sigmoid activation function for compression of each pixel to a value between zero and one.

\subsubsection{Shadow Removal Network}
\begin{figure}[H]
    \centering
    \includegraphics[width=\textwidth,height=\textheight,keepaspectratio]{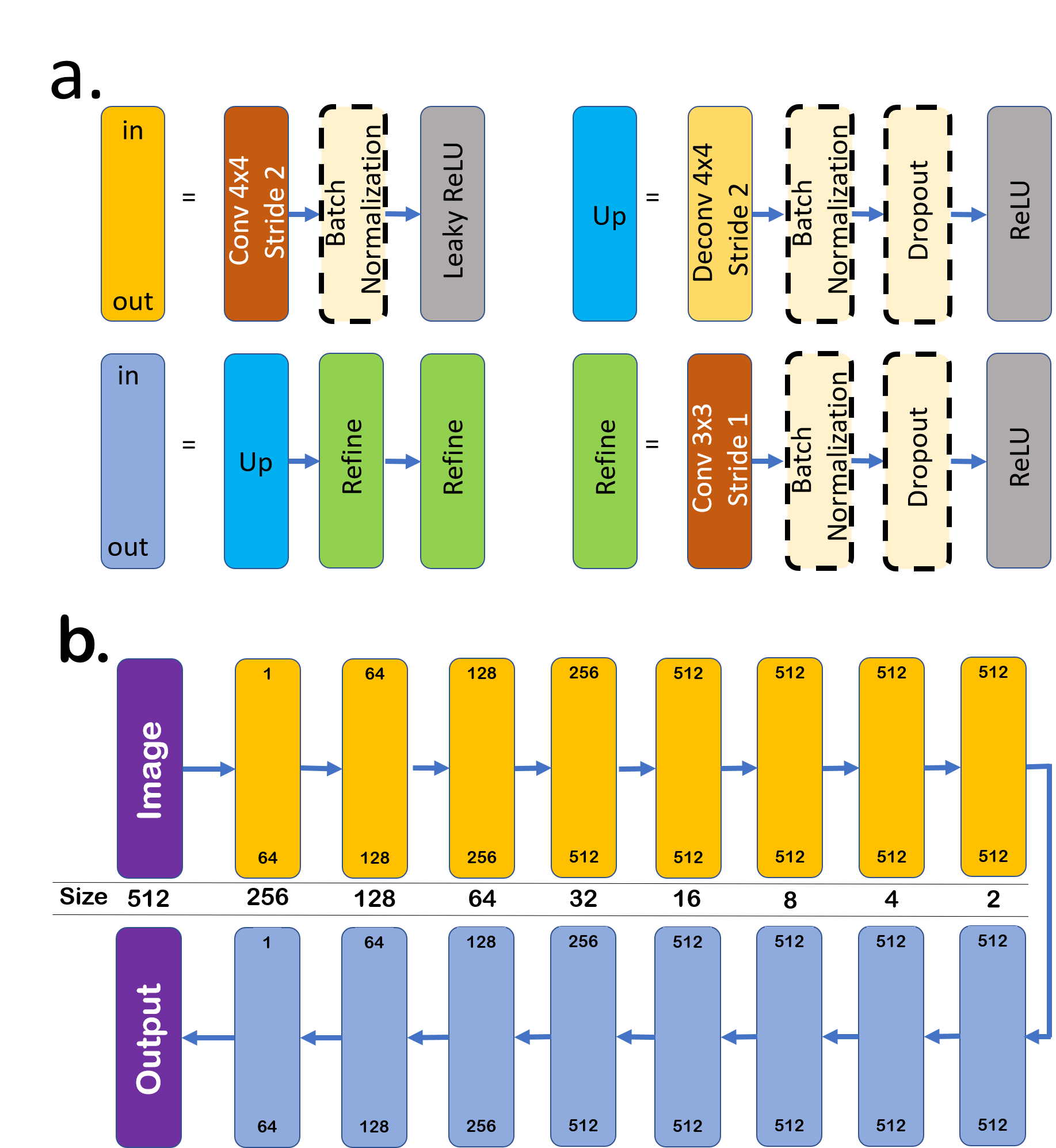}
    \caption{All arrows represent a forward pass of the output from one layer to the input of the next layer. Each box represents a module (a set of layers). The size of our input image is $512 \times 512$. \textbf{(a)} Definitions of the layers in downsampling and upsampling modules within the shadow removal network. Dotted boundaries indicate that the module is present only within some layers. In and out values at the top and bottom of each rectangle represents the number of feature maps being input and output from that module respectively. \textbf{(b)} Size row is the size of the output of each module (rectangles above and below it).}
    \label{fig:3}
\end{figure}

The shadow removal network was inspired from the Deep Video Portraits approach of \cite{RN15}. A schematic of the architecture is shown in Figure \ref{fig:3}. Baseline images were inputted into the network and passed through a downsampling segment and an upsampling segment (colored in yellow and blue, respectively; Figure \ref{fig:3}b). The downsampling segment allowed the network to understand contextual information, while the upsampling segment increased the resolution of the output. Features from both segments were combined to produce a more precise output in the successive convolution layer.\cite{RN78}
The shadow removal network consisted of 8 downsampling modules and 8 upsampling modules (Figure \ref{fig:3}b). The first encoding layer and the last decoding layer did not employ batch normalization. We included a dropout of $0.5$ only in the first $3$ upsampling layers. The network had $55.7$M parameters, and occupied $820$ Mo of RAM on a Nvidia GTX 1080Ti GPU. Each module in the downsampling module consisted of a convolution layer (stride $2$, kernel size $4 \times 4$) followed by a batch normalization and a leaky ReLU activation function. Every downsampling module reduced the feature map size by half, enabling the network to derive contextual information and encode its input into an increasing number of filters. The maximum number of filters plateaued at $512$ for $4$ times, and then moved on to the decoding segment.\\

Each decoding segment consisted of $3$ other submodules, namely, Up, and two Refine submodules. The Up submodule consisted of a transpose convolution (size $4 \times 4$, stride $2$) followed by a batch normalization and a ReLU activation function. Every Up submodule allowed the network to improve its decoding efficiency from encoded information from the input. The Refine submodule consisted of a convolution (size $3 \times 3$, stride $1$) followed by a batch normalization and a 0.5 dropout. We repeated this process until a $512 \times 512$ feature map was obtained and we reduced the number of feature maps produced in the last layer to 1 to mimic input images. Finally, we applied a pixel-wise sigmoid activation to compress all activations from the decoding segment to values between 0 and 1.

\subsection{Image Augmentation}

An image augmentation network was created using Pytorch \cite{RN11} to perform on demand image augmentation during training. Our data augmentation consisted of random transformations including: horizontal flipping, image rotations (angle between $-\ang{40}$ and $\ang{40}$), XY translations ($-20\%$ to $20\%$ of the image size), image scaling (scaling factor between $0.8$ and $1.2$), and image shear (shear angle between $-\ang{20}$ and $\ang{20}$). All images were then resized to $512 \times 512$ pixels. 

\subsection{Weighted-mixture loss function for the shadow removal network}

An adversarial shadow removal network was simultaneously trained using a custom loss function (to be minimized during training) that reduced the appearance of shadows in output images. This custom loss function was used to restore structural information under blood vessel shadows while maintaining structural information in all other areas. It consisted of a tuned weighted combination of four losses: the content, style, total variation, and shadow losses, as briefly explained below.\\

\begin{figure}[H]
    \centering
    \includegraphics[width=\textwidth,height=\textheight,keepaspectratio]{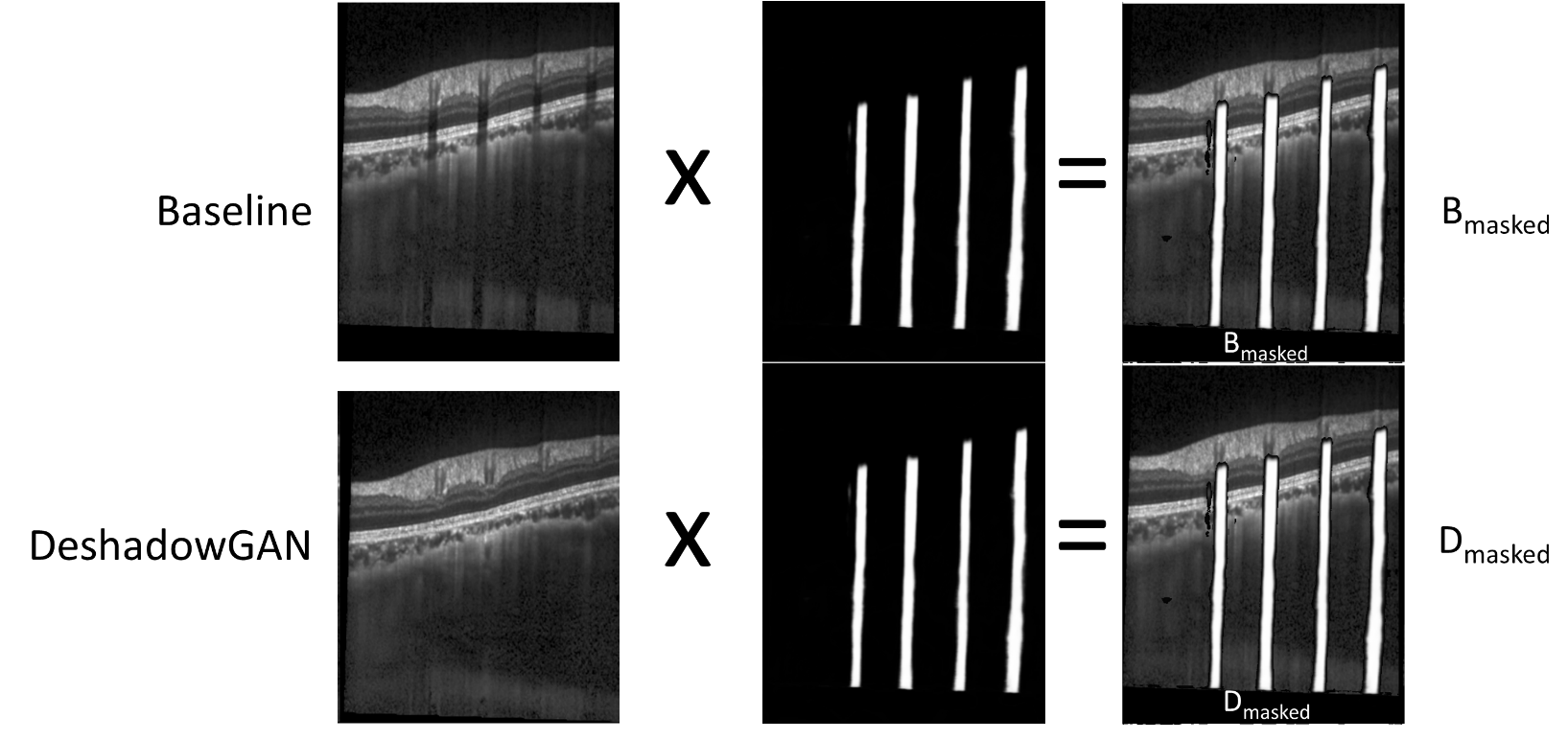}
    \caption{Masking of baseline and deshadowed images during content loss and style loss calculations. Predicted shadow mask for the baseline image is used to mask both the baseline and deshadowed image.}
    \label{fig:4}
\end{figure}

\textbf{Content Loss.} We used the content loss to ensure that all non-shadowed regions of a given image remained the same after shadow correction. To do so, we compared the high-level image feature representations (using a pretrained Resnet152 architecture) from a given baseline image with those from its deshadowed output. Note that the content loss function has been used in Style Transfer \cite{RN16}  and has been shown to maintain fine image details and edges. 
To calculate the content loss, we first segmented all shadows from the baseline image with our shadow detection network. All shadows were then masked (pixel intensity values equal to zero) in order to generate two images: $B_{\textrm{masked}}$ (baseline image with masked shadows) and $D_{\textrm{masked}}$ (deshadowed image with masked shadows) as shown in Figure \ref{fig:4}. 

$B_{\textrm{masked}}$ and $D_{\textrm{masked}}$ were then passed to the Resnet152 network \cite{RN72} itself trained using the ImageNet dataset. The content loss was then calculated for each image pair as an Euclidean norm as follows:

\begin{equation} \label{eq:1}
\mathcal{L}_{\textrm{content}} (B_{\textrm{masked}},D_{\textrm{masked}})= \sum\limits_{i \in \{9, 33, 141\}}\frac{1}{C_{i} H_{i} W_{i}}  \lvert P_{i} (B_{\textrm{masked}} )-P_{i} (D_{\textrm{masked}} )\rvert^{2}
\end{equation}

where $P_i(x)$ is a feature map that contains the activations of the $i$-th convolutional layer of the Resnet152 network for a given image; the quantities $C_i$, $H_i$ and $W_i$ represent the channel number, height, and width of the feature map $P_i(x)$, respectively. The convolutional layers mentioned in equation \eqref{eq:1} (i.e. $9$, $33$, $141$, and $150$) were selected because they are the final convolutional layers in the Resnet152 network before downsampling.\\ 

\textbf{Style Loss.} On top of the content loss, we also used the style loss \cite{RN16} to ensure that the image's style (texture) remained the same in the non-shadowed regions after shadow correction. We compute the gram matrix of an image to find a representation of its style. The style loss was then computed for each image pair ($B_{\textrm{masked}}$, $D_{\textrm{masked}}$) and was defined as the Euclidean norm between the Gram matrices of $B_{\textrm{masked}}$ and $D_{\textrm{masked}}$:
\begin{equation} \label{eq:2}
	\mathcal{L}_{\textrm{style}} (B_{\textrm{masked}},D_{\textrm{masked}})= \sum\limits_{i \in \{9, 33, 141\}} = \left| G_i(B_{\textrm{masked}})-G_i (D_{\textrm{masked}})\right|^2
\end{equation}
where $G_i$ is a $C_i\times C_i$  matrix defined as $G_i(x)= P_i(x)_{C_i,W_i,H_i} \times P_i(x)_{H_i,W_i,C_i}$.\\

\textbf{Total variation loss.} We used the total variation loss to prevent checkerboard artifacts from appearing in deshadowed images. It was defined as the sum of the differences between neighboring pixels in a given deshadowed image, D:
\begin{equation}\label{eq:4}
	\mathcal{L}_{\textrm{TV}}(D)=  \frac{1}{n}  \sum\limits_{i,j}\lvert D_{i+1,j}-D_{i,j} \rvert + \lvert D_{i,j+1}-D_{i,j} \rvert
\end{equation}
where $n$ is the total number of pixels in the deshadowed image and $i$ and $j$ are the row and column numbers, respectively.\\

\textbf{Shadow Loss.} The shadow loss was defined to ensure that shadows were properly removed so that they become undetectable to the shadow detection network. Once a given image D was deshadowed, it was passed to the shadow detection network to produce a predicted shadow mask, $M_{D}$ (with pixel intensities equal to 1). All pixel intensities in the shadow mask were summed and this sum was defined as the shadow loss function. 	\\

\textbf{Total loss.} The total loss for the shadow removal network was defined as
\begin{equation}\label{eq:5}
\mathcal{L}_{\textrm{total}} = 
w_1 \times (\textrm{content loss}) + 
w_2 \times (\textrm{style loss}) + 
w_3 \times (\textrm{shadow loss}) + 
w_4 \times (\textrm{total variation loss})
\end{equation}
where $w_1$, $w_2$, $w_3$ and $w_4$ are weights that were given the following values: $100$, $0.1$, $100$, and $1 \times 10^{-5}$, respectively. Note that all weights were tuned manually through an iterative approach. First, we found that with a value of $w_1=100$, we generated images with no content loss (i.e. no structural changes), but not without the presence of checkerboard artifacts. These artifacts could be removed when choosing $w_2=0.1$ and $w_4=1e-5$.[28] Finally, we increased the value of $w_3$ until the shadow loss became the largest component in the total loss function (so that the focus remained on removing shadows), and until shadow removal was deemed qualitatively acceptable for smaller width shadows. This was optimum when $w_3=100$. 

\subsection{Training Parameters}
All training and testing were performed on a Nvidia GTX 1080 Ti GPU with CUDA 10.1 and cuDNN v7.6.0 acceleration. Using these hardware specifications, each image took an average of $10.3$ ms to be deshadowed. Training was performed using the Adam optimizer at a learning rate of $1\times 10^{-5}$ and a minibatch size of $b=2$. A learning rate decay was implemented to halve learning rates every $10$ epochs. We stopped the training when no observable improvements in output images could be observed.

\subsection{Shadow Removal Metrics}
\textbf{Intralayer Contrast.} We used the intralayer contrast to assess the performance of our algorithm in removing shadows. The Intralayer contrast was defined as
\begin{equation}\label{eq:6}
\textrm{Intralayer Contrast} = \left| \frac{I_1-I_2}{I_1+I_2} \right| 
\end{equation}
where $I_1$ is the mean pixel intensity from 5 manually-selected regions of interest (size: $5\times 5$ pixels) that are shadow free in a given retinal layer, and $I_2$ is that from 5 neighboring shadowed regions of the same tissue layer. The Intralayer contrast varies between 0 and 1 where values close to 0 indicate the absence of blood vessel shadows, and values close to 1 indicating the strong presence of blood vessel shadows. 
We computed the intralayer contrast for multiple tissue layers of the ONH region, namely the RNFL, the PR, IPL and the RPE, before and after application of our deshadowing algorithm. The intralayer contrast was computed on an independent test set consisting of $291$ images. Results were reported in the form of mean $\pm$ standard deviation. 

\subsection{Comparison with Adaptive Compensation}
To evaluate the effectiveness of our deshadowing algorithm, we compared images deshadowed using DeshadowGAN with images enhanced with adaptive compensation \cite{RN6, RN73}, the gold-standard for correcting OCT shadows. For adaptive compensation, we used contrast, decompression, compression, and threshold exponents of $1$,$4$,$4$, and $6$, respectively. Intralayer contrasts were also computed for all compensated images (same regions as those used for the baseline images). 

\subsection{Validation using a Test Scenario with Known Ground Truth}
We investigated whether our deshadowing algorithm was capable of restoring information below blood vessel shadows without introducing unwanted artifacts. To do so, we require ground-truth images without blood vessel shadows but such images are not easy to obtain in vivo. An alternative is to add artificial shadows to a given baseline image and assess whether our algorithm can remove them without introducing artifacts.  \\

Accordingly, we created exponential decay maps on images to simulate the effect of light attenuation and trained DeshadowGAN with such images. A shadow can be simply simulated as:
\begin{equation}\label{eq:7}
{(\textrm{Shadow Pixel})}_{ij}={(\textrm{Baseline Pixel})}_{ij} \times e^{-\alpha i}
\end{equation}
where $i$ is the row number and $\alpha$ indicates the rate of decay. We used the same training and testing image sets, except that 2 artificial shadows (random width between $1$ and $100$ pixels; random $\alpha$ between $100$ and $300$) were randomly added to each baseline image. DeshadowGAN was re-trained with the exact same aforementioned procedure, including the manual segmentation of the artificial shadows. Note also, that during training, the DeshadowGAN algorithm did not have access to the ground truth baseline images without shadows. After deshadowing, the presence of artifacts was assessed qualitatively. 

\section{Results}
\label{sec:results}
\subsection{DeshadowGAN Decreased the Intralayer Contrast}
\begin{figure}[H]
    \centering
    \includegraphics[width=\textwidth,height=\textheight,keepaspectratio]{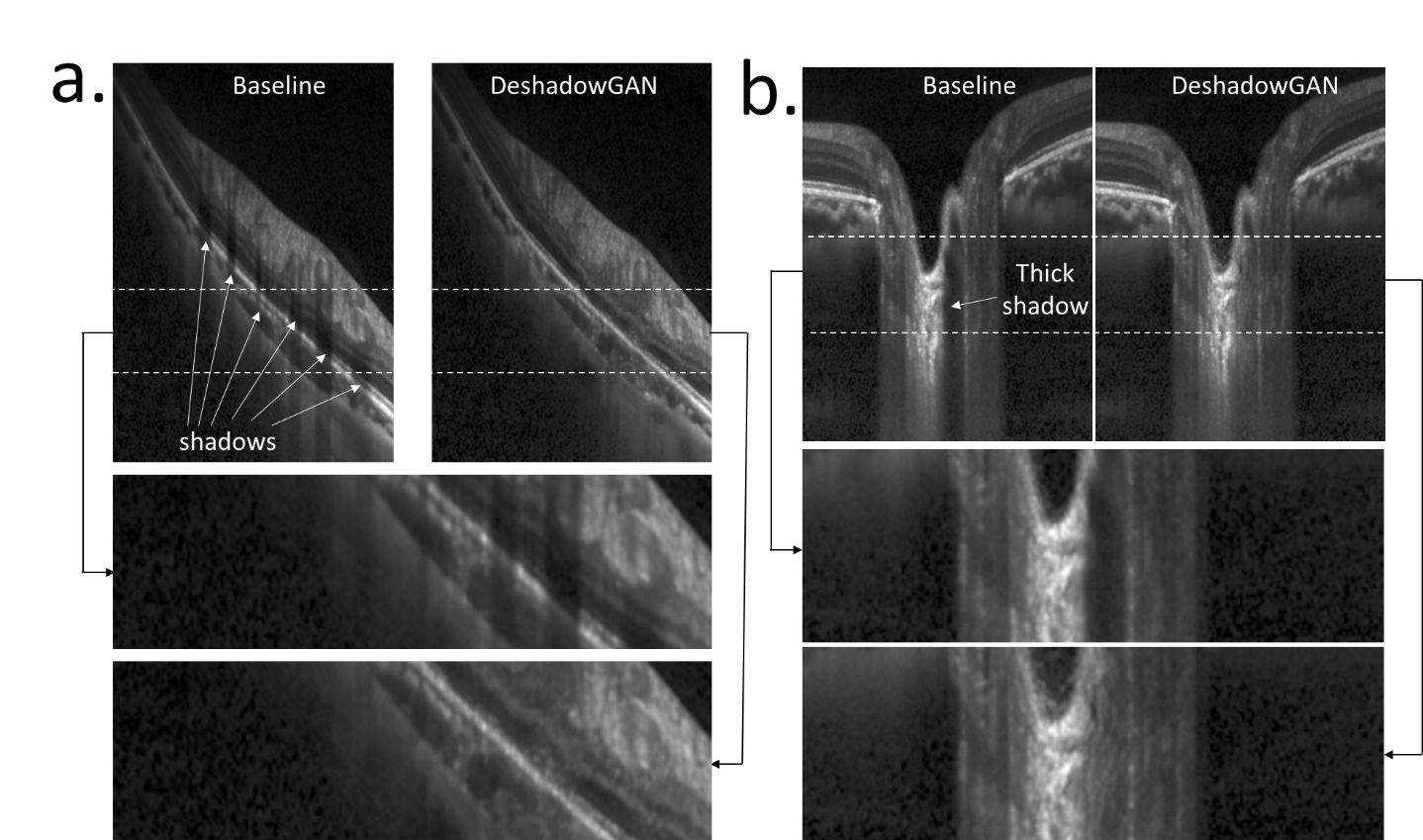}
    \caption{Images of retinal layers before and after deshadowing of areas \textbf{(a)} away from the optic disc, and \textbf{(b)} areas around the optic disc.}
    \label{fig:5}
\end{figure}

After application of our algorithm, blood vessel shadows from unseen images were successfully identified and corrected from each retinal layer as observed quantitatively and qualitatively. On average, we observed improvements in intralayer contrast of $33.7\pm 6.81\%$, $28.8\pm 10.4\%$, $35.9\pm 13.0\%$, and $43.0\pm 19.5\%$ for the RNFL, IPL, PR, and RPE layers respectively. This can be qualitatively observed in a B-Scan of the peripapillary tissues in Figure \ref{fig:5}.

\subsection{Comparison with Adaptive Compensation}
\begin{figure}[H]
    \centering
    \includegraphics[width=\textwidth,height=\textheight,keepaspectratio]{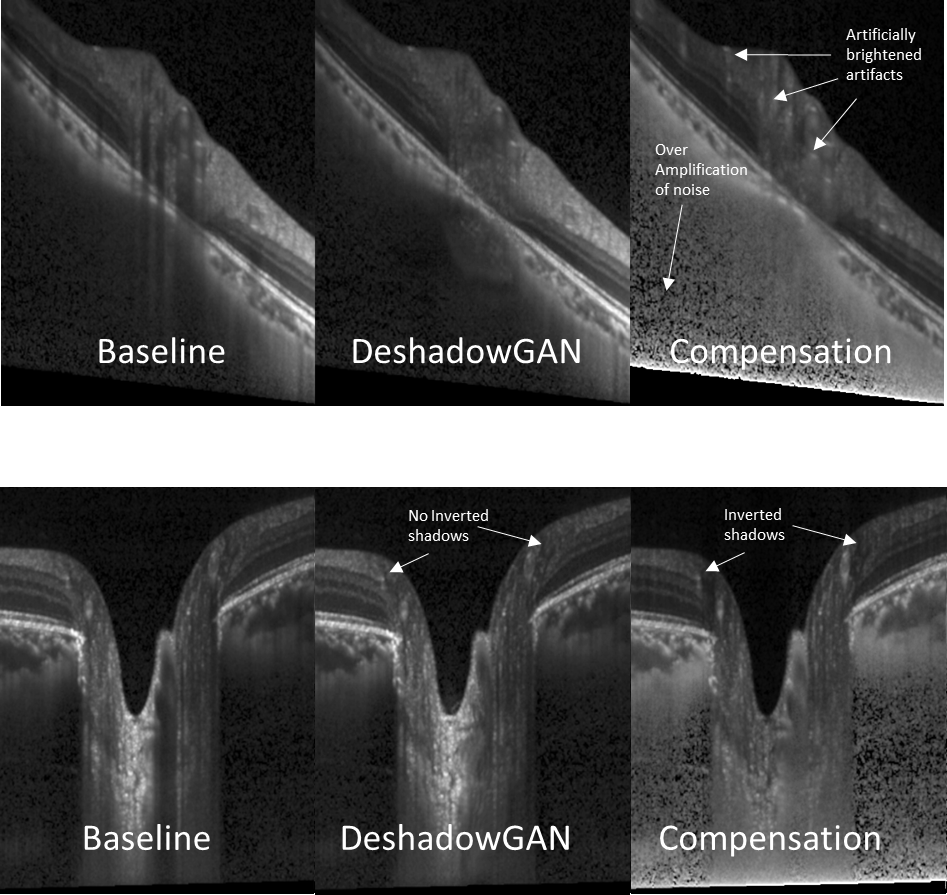}
    \caption{Compensation Artifacts Comparison with DeShadowGAN. Artificially brightened artifacts and over-amplification of noise in compensated image (top right). Inverted shadows in compensated images (bottom right).}
    \label{fig:6}
\end{figure}

DeshadowGAN was able to correct shadows without affecting the contrast of anterior layers, without adding noise, and without creating artifacts (Figure \ref{fig:6}).

\begin{figure}[H]
    \centering
    \includegraphics[width=\textwidth,height=\textheight,keepaspectratio]{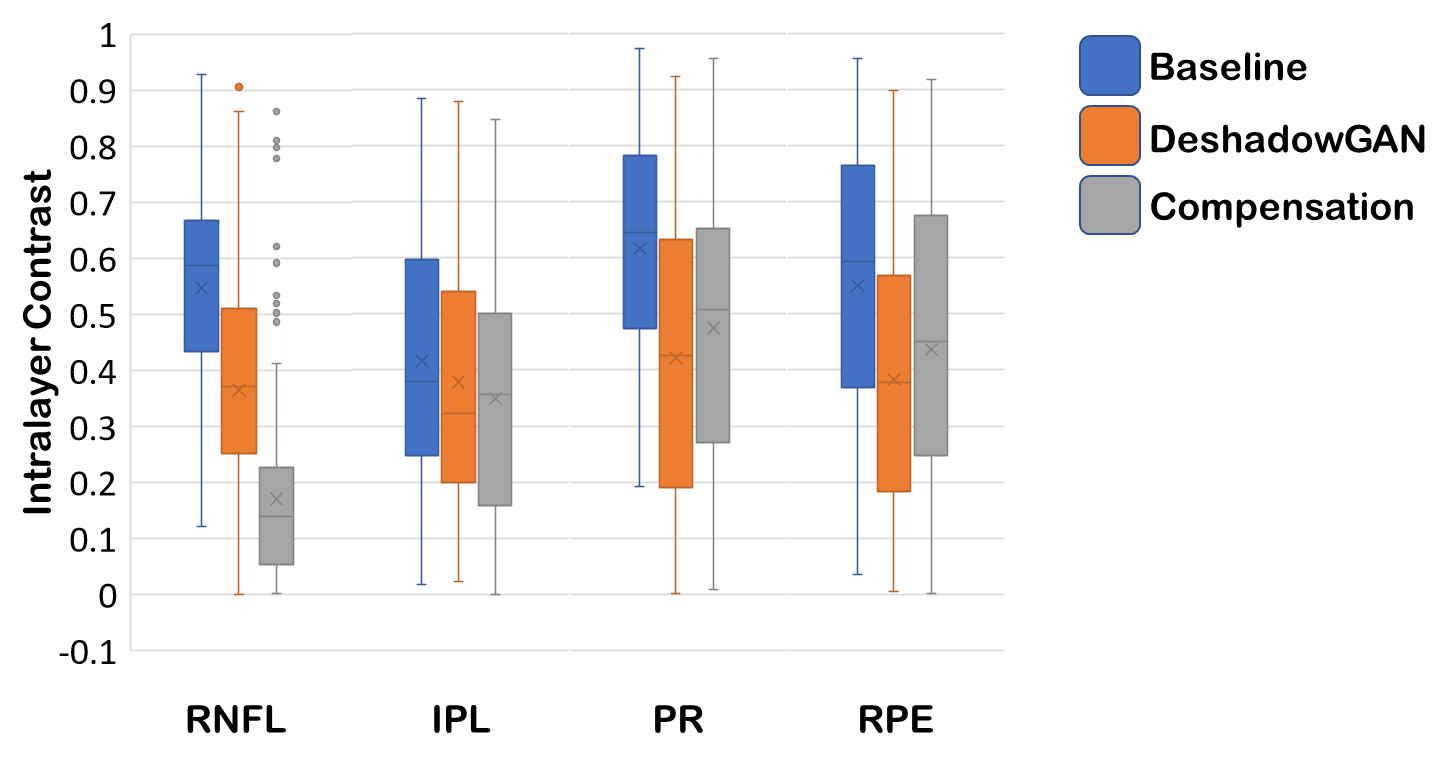}
    \caption{Intralayer contrast comparison between baseline, deshadowed and compensated images. When compared with compensation, DeShadowGAN tends to perform better in deeper layers.}
    \label{fig:7}
\end{figure}

In addition, DeshadowGAN had better shadow removal capabilities than compensation as layer depth increased. This can be observed from the box plot in Figure \ref{fig:7}, where the $25$-th and $75$-th percentiles of the intralayer contrast for DeshadowGAN gradually increased against those of compensation from the RNFL to the RPE layers. 

\begin{figure}[H]
    \centering
    \includegraphics[width=\textwidth,height=\textheight,keepaspectratio]{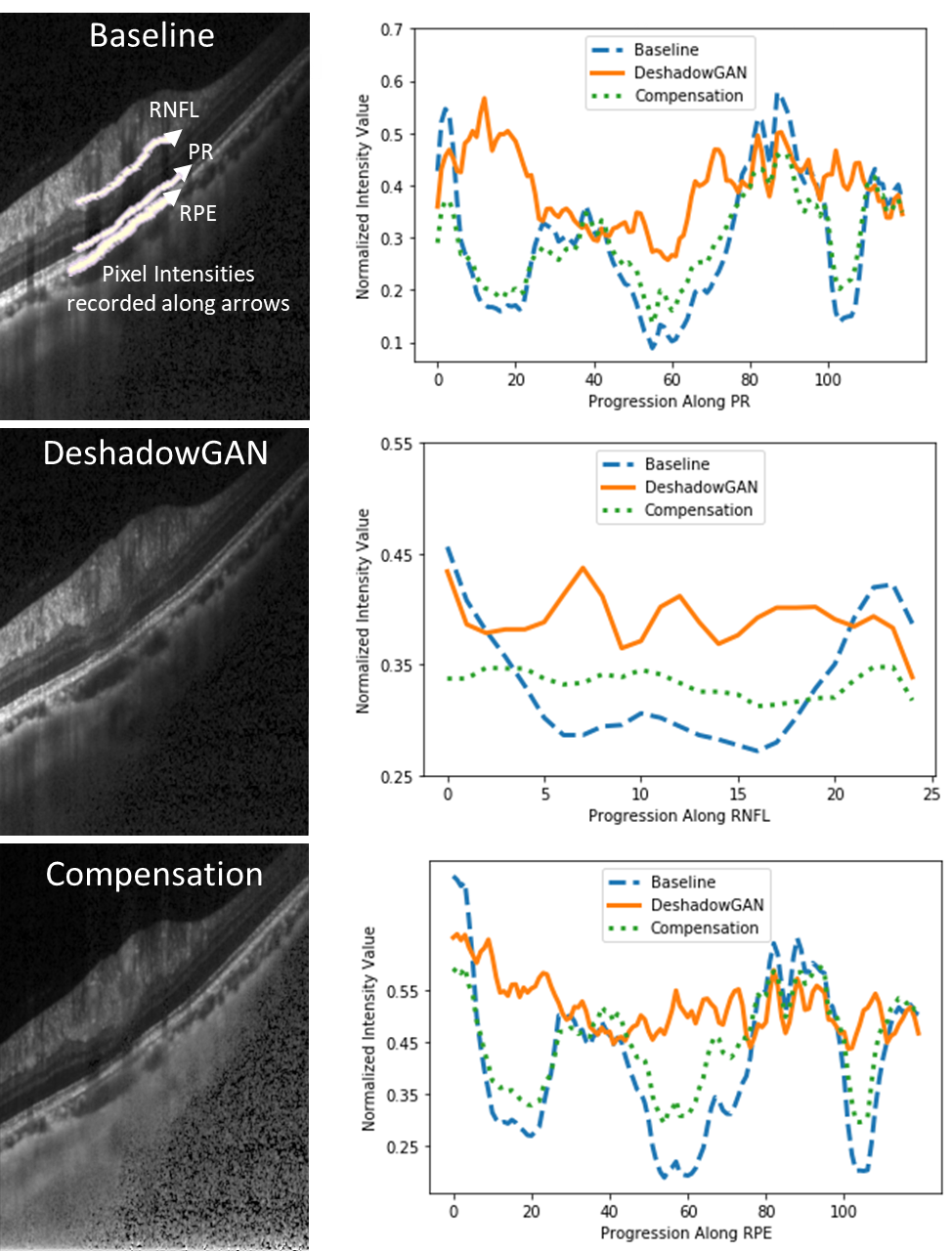}
    \caption{Layer-wise lateral pixel intensities across the PR, RPE, and RNFL layers. Direction of progression is along arrow at the bottom of each image.}
    \label{fig:8}
\end{figure}

Shadow removal was also qualitatively corroborated by observation of the flattened lateral pixel intensities (across shadows) for the PR, RPE, and RNFL layers before and after shadow removal (Figure \ref{fig:8} -- right column). DeshadowGAN recovered the shadows to a larger extent as compared to compensation. Furthermore, we observed that compensation did not have an increase in shadow information but rather a decrease in non-shadow intensities in shallow layers, as non-shadow pixel intensities were found to be up to $50\%$ lower after compensation.

\subsection{Proof of Principle: DeshadowGAN Did Not Create Artifacts}
\begin{figure}[H]
    \centering
    \includegraphics[width=\textwidth,height=\textheight,keepaspectratio]{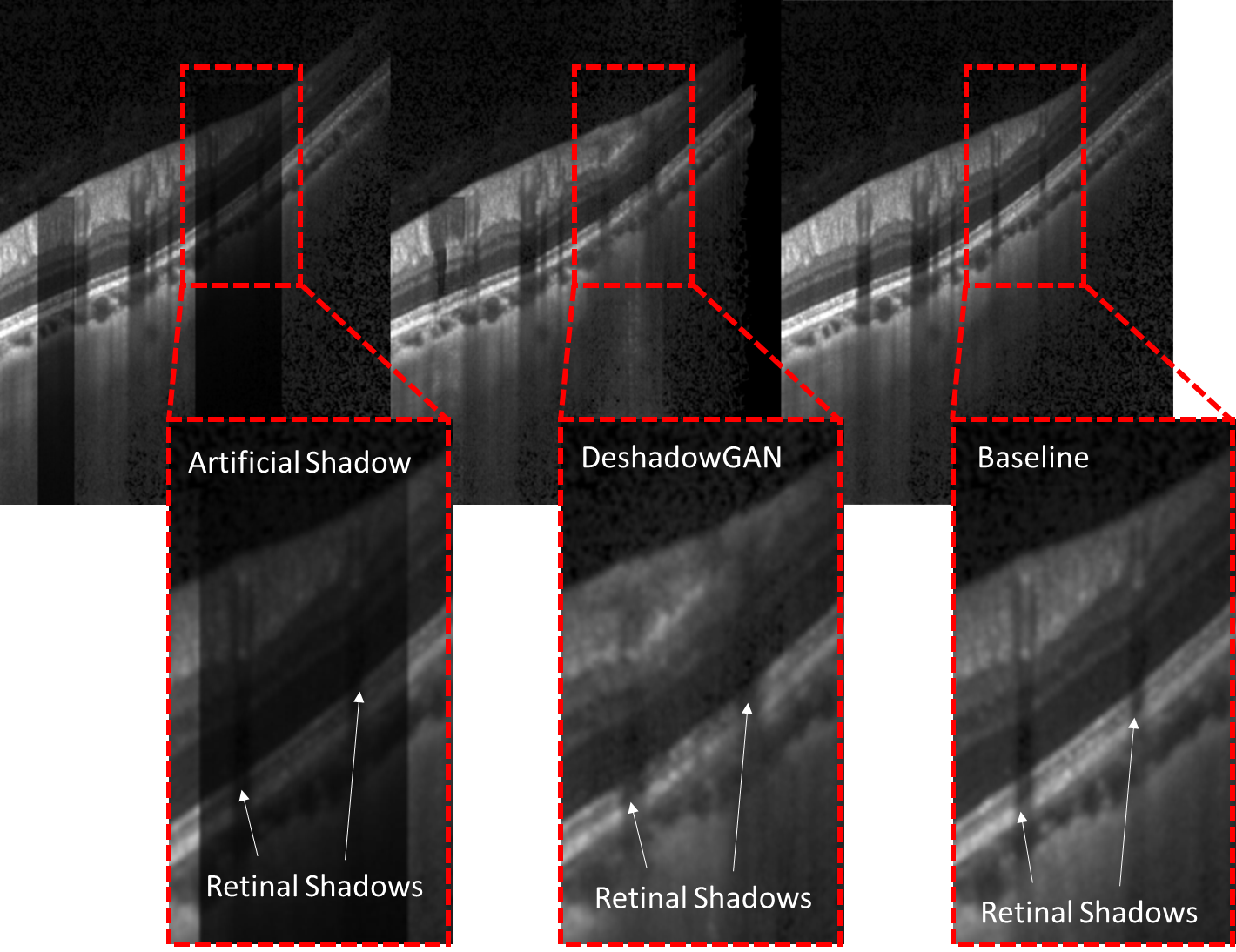}
    \caption{Artificial shadow removal experiment results. From left: Baseline with artificial shadow, deshadowed image from DeshadowGAN, Baseline image without artificial shadow.}
    \label{fig:9}
\end{figure}

Qualitative analysis of our results showed that no artificial anatomical information was created within deshadowed images. This can be qualitatively observed from Figure \ref{fig:9}, where both genuine retinal shadows were retained, albeit not as clearly defined as compared to the ground truth (baseline images in this case). 

\section{Discussion}
\label{sec:discussion}
In this study, we have proposed a novel deep learning algorithm (DeshadowGAN) with a weighted-mixture loss function to remove retinal blood vessel shadows in OCT images of the ONH. When trained with baseline OCT images and manually created binary shadow masks, DeshadowGAN improved tissue visibility under shadows at all depth, regardless of shadow width. DeshadowGAN may be considered as pre-processing to improve the performance of a wide range of algorithms including those currently being used for OCT image segmentation, denoising, and classification.\\

Having successfully trained, validated and tested our algorithm with a total of $2,619$ baseline OCT images, we found that DeshadowGAN can be applied to new images not previously seen by the network in order to correct shadows. Furthermore, for new images, DeshadowGAN does not require any segmentation, delineation or identification of shadows by the user. Our results confirmed consistently higher intralayer contrasts, flatter layer-wise pixel intensity profiles across shadows, and the absence of many artifacts commonly found in compensated images. Thus, we may be able to provide a robust deep learning framework to consistently remove retinal blood vessel shadows of varying sizes and intensities. \\

In addition, DeshdowGAN was able to successfully eliminate the deleterious effects of light attenuation affecting the visibility of retinal layers and deeper tissues such as the LC. DeshadowGAN helped substantially recover the visibility of the anterior lamina cribrosa (ALC) boundary, where sensitive pathophysiologic deformation could signal the onset of early glaucoma.\cite{RN17, RN88, RN89} Deep collagenous tissues such as the LC and adjacent peripapillary sclera (PS) are the main load-bearing tissues of the eye in the ONH region,\cite{RN6} and it has been reported that biomechanical and morphological changes in these tissues may serve as risk factors for glaucoma.\cite{RN91, RN92, RN93} The robustness of the OCT-based measurements performed on these tissues could be substantially improved after application of our proposed algorithm.\\

Corrected images with DeshadowGAN did not exhibit strong artifacts that can often be observed with adaptive compensation such as: inverted shadows, hyper-reflective spots, noise over-amplification at high depth (see examples in Figure \ref{fig:6}), and hypo-reflective retinal layers. For this latter case, we found that compensation can indeed reduce tissue brightness in the anterior retinal layers (while enhancing deeper connective tissue layers) by up to 50\%. Brightness is typically not affected with DeshadowGAN. We also believe that compensation artifacts could cause issues for automated segmentation algorithms that rely on the presence of homogenous pixel intensity values within the same layer. \cite{RN84, RN85, RN86} Because DeshadowGAN generates significantly less artifacts, it has the potential to be used as an AI pre-processing step for many automated OCT applications in ophthalmology, such as, but not limited to: segmentation, denoising, signal averaging, and disease classification \cite{RN99, RN100, RN101, RN102, RN103}.   \\

As a first proof-of-principle, we also found that DeshadowGAN did not create anatomically inaccurate information under shadows while maintaining all other image regions true to their original quality. However, this was only confirmed with artificial data by simply adding fake shadows (simulated as an exponential decay). If one wanted to confirm such results with ex- or in-vivo data, one would need to image the exact same tissue region with and without the presence of blood flow. Such experiments would be extremely complex to perform, especially in humans in vivo, even if blood is flushed with saline temporarily (as is done with intravascular OCT). However, we understand that such validations may be necessary for full clinical acceptance of this methodology. From our point of view, it would also be imperative to further confirm that DeshadowGAN would not interfere with another AI algorithm aimed at improving diagnosis or prognosis. On the other hand, it also very possible that DeshadowGAN may increase diagnosis/prognosis performance of other algorithms, and we hope to test such hypotheses in details in the future.\\  

Several limitations of this work warrant further discussion. While DeshadowGAN has performed relatively well on baseline OCT images from healthy eyes, we cannot confirm that its performance will remain the same for eyes with pathological conditions such as glaucoma. This is because deep learning approaches respond unpredictably when the input is very different from its training images,\cite{RN94, RN95} and `pathological' training sets may be required. Furthermore, DeshadowGAN was trained on high quality multi-frame OCT images from a single Spectralis OCT device. It is unknown if the algorithm would be able to perform as effectively if applied to OCT images obtained from other OCT devices, or OCT images from the same device but with significantly less or no signal averaging. Similarly, each scenario may require a separate training set. We aim to perform further tests to assess all possible scenarios. \\

In conclusion, we propose a novel algorithm to correct blood vessel shadows in OCT images. Such an algorithm can be considered as a pre-processing step to improve the performance of a wide range of algorithms including those currently being used for OCT image segmentation, denoising, and classification.  \\

\section*{Acknowledgements}
Singapore Ministry of Education Academic Research Funds Tier 1 (R-155-000-168-112 [AHT];R-397-000-294-114 [MJAG]); National University of Singapore (NUS) Young Investigator Award Grant (NUSYIA FY16 P16;R-155-000-180-133; AHT); National University of Singapore Young Investigator Award Grant (NUSYIA FY13 P03;R-397-000-174-133 [MJAG]); Singapore Ministry of Education Academic Research Funds Tier 2 (R-397-000-280-112 [MJAG] and MOE2016-T2-2-135 [AHT]);National Medical Research Council (Grant NMRC/STAR/0023/2014 [TA]).

%\section*{Disclosures}
%MG and AT are the co-founders of Abyss Processing Pte Ltd. 

\bibliographystyle{unsrt}
\bibliography{deshadowgan}

\end{document}